\documentclass[conference]{IEEEtran}
\IEEEoverridecommandlockouts
\usepackage{cite}
\usepackage{hyperref}
\usepackage{amsmath,amssymb,amsfonts}
\usepackage{algorithmic}
\usepackage{graphicx}
\usepackage{textcomp}
\usepackage{xcolor}
\def\BibTeX{{\rm B\kern-.05em{\sc i\kern-.025em b}\kern-.08em
    T\kern-.1667em\lower.7ex\hbox{E}\kern-.125emX}}
    
\usepackage[T1]{fontenc}
\usepackage[numbered,framed]{matlab-prettifier}
\begin{document}

\title{SiGe BiCMOS Circuit Design using only PMOS and HBTs Approach for the Ocean Worlds Exploration}
%%for Extreme Cold Temperatures and High Radiation Environments

\author{
\IEEEauthorblockN{Md Omar Faruk\IEEEauthorrefmark{1}, Steven Corum\IEEEauthorrefmark{1}, Zakaraya Hamdan\IEEEauthorrefmark{1}\IEEEauthorrefmark{2}, Alex Seaver\IEEEauthorrefmark{1},  Travis Graham\IEEEauthorrefmark{1}\IEEEauthorrefmark{2},  Benjamin J. Blalock\IEEEauthorrefmark{1}}
%\\Jeffrey W Teng\IEEEauthorrefmark{3}, John D Cressler\IEEEauthorrefmark{3}}

%Linda Del Castillo\IEEEauthorrefmark{3}, Mohammad M. Mojarradi\IEEEauthorrefmark{3}
\IEEEauthorblockA{\IEEEauthorrefmark{1}Dept. of Electrical Engineering and Computer Science,
University of Tennessee, Knoxville, TN, USA. }
\IEEEauthorblockA{\IEEEauthorrefmark{2}Texas Instruments, Knoxville, TN, USA.}
%\IEEEauthorblockA{\IEEEauthorrefmark{3}School of Electrical and Computer Engineering, Georgia Tech, Atlanta, GA, USA.}
%\IEEEauthorblockA{\IEEEauthorrefmark{4}Jet Propulsion Laboratory, Caltech, Pasadena, CA, USA.}
Email: mfaruk@vols.utk.edu
}

\maketitle
\thispagestyle{plain}
\pagestyle{plain}
\begin{abstract}
Space exploration to have the biosignatures of extraterrestrial life on different planets with oceans in our solar system and beyond requires the design and manufacturing of robust and reliable electronic systems that can be used for sensing, data processing, controlling motor/actuators, and communication while surviving an extreme environment. Commercial-off-the-shelf (COTS) components cannot survive a long time in such harsh environments after being housed in a ``Warm-Electronics-Box'', and any electronic system designed for such extreme conditions must be tailored to suit such operation. The presence of extremely cold temperatures and high radiation adversely affects the device parameters over time, i.e. the operation of electronic systems.   
\end{abstract}

\begin{IEEEkeywords}
SiGe BiCOMS, PMOS, HBTs, Ocean Worlds exploration.
\end{IEEEkeywords}

\section{Introduction}\label{sec:section_guide}

The tell of life has an intimate connection with the oceans just like on our home planet, the Earth, where the oceans take about 70$\%$ of the surface and drive a water cycle that helps to sustain life. But if someone zooms out of Earth and looks into the void of the universe, the story of oceans on Earth puts us in a large family of ocean worlds that goes beyond our solar system \cite{Nasaow}. In the 2016 budget, Congress funded NASA to establish the ``Ocean Worlds Exploration Program (OWEP)'' to probe into the worlds with oceans in our outer solar system that may have liquid water beneath the ice shelf to check future habitability and collect samples as the biosignatures of extraterrestrial life. Europa, Enceladus, and Titan are mentioned in the 2016 House Appropriations bill as ocean worlds, but the Galileo and Cassini spacecraft have some geophysical measurements that prove that there are subsurface oceans in Ganymede and Callisto, which were later included in the ocean worlds program \cite{Roadmap}. 
Among the worlds with subsurface oceans, Europa and Enceladus are prioritized due to evidence of connectivity between the ocean and the surface (approximately 10 km and 30 km) and probable habitability due to the interaction between the oceans and the seafloor with rocks \cite{Europafacts, Enceladusafacts}. The surfaces of Titan, Ganymede, and Callisto are blanketed with very thick (approximately 55 km, 100 km, and 250 km, respectively) ice shells, which makes access to the liquid water oceans very difficult \cite{Titanfacts, Ganymedefacts, Callistofacts}. The Jupiter Icy Moons Explorer (Juice) was recently sent by the European Space Agency (ESA) to study Jupiter's three Galilean moons, i.e., Europa, Ganymede, and Callisto \cite{Roadmap}. It will reach Jupiter in July 2031. NASA is implementing the Europa Clipper orbiter mission under the ``Ocean Worlds Exploration'' program to Europa at the end of 2024 \cite{Europaclipper} and a lander shortly after \cite{Europalander}.

\begin{figure} 
         \centering 
         \includegraphics[scale=0.73]{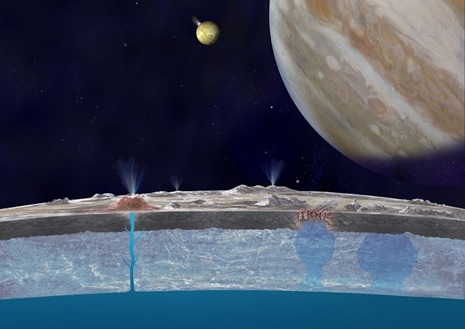}
         \caption[ Europa’s interior Ocean (An artist’s view). ]
         { Europa’s interior Ocean (An artist’s view) \cite{Europa}.}
         \label{fig:Europa}
\end{figure}
%\clearpage}

 Environmentally invariant electronic systems needed for computing, sensing, controlling, and communications are fundamental for NASA's ``Concepts for Ocean Worlds Life Detection Technology (COLDTech)'' program, which is designed to explore the oceans with liquid water in Europa and two other moons of Jupiter under their thick (10 km for Europa) ice shelf to check future habitability and collect samples as the biosignatures of extraterrestrial life. But the surface (or below) environment in these Ocean Worlds is very harsh for any COTS (Commercial off-the-shelf) electronic system with extremely high radiations (total ionizing dose (TID) of  5 Mrad(Si)) and extremely low temperatures  (93K = -180°C ). For NASA's Mars rover, all the electronics are placed in ``shielded warm electronics vaults'' to make it work within the mission parameters  \cite{YChen}, but for Ocean Worlds, where the need for electronics is very broad,  from battery-operated surface operation where the environment is the harshest (constant exposure to ionizing radiation and cryogenic temperature), then operation during the drilling of the ice shelf up to the ocean, and finally the exploration in the somewhat friendlier ocean with the submersible. So, for any mission to Jupiter ocean worlds, environmentally invariant electronics (i.e. robust operation regardless of the environment) are very desirable from a mission-science perspective so that they can work not only on the harsh surface environment but also in the somewhat friendlier below-surface missions. So, to improve mission science, these envisioned environmentally invariant electronics platforms need not only to allow digital/analog/RF circuit functions but also to be available commercially, highly integrated, minimize cost and introduce comprehensive size-weight-power-cost benefits \cite{COLDTech}. Due to this increased interest in space exploration, these environmentally invariant electronics have to be radiation-hardened (rad-hard) and cold-capable. For terrestrial applications, CMOS offers the best solution in terms of power, performance and area (PPA), but for extraterrestrial applications, CMOS electronics have huge reliability issues, such as increased leakage current of off-state NMOS devices and single event latch-up in a radiating environment below a total Mrad ionizing dose (TID) \cite{Zakthesis, Cressler}. Furthermore, at cryogenic temperatures, the generation of hot carriers in NMOS deteriorates the situation even further by reducing the lifetime by one or two orders of magnitude shorter than that of PMOS \cite{Clark, YChen}. However, PMOS transistors provide performance that is relatively better than that of NMOS transistors in both cases. Due to these issues with NMOS, an alternative technology route is needed to CMOS. The Silicon Germanium (SiGe) Heterojunction Bipolar Transistor (HBT) is an alternative solution to design circuits that are rad-hard and cold-capable, due to their tolerance of multi-Mrad TID and temperatures as low as 1 K \cite{Cressler}. So, BiCMOS is an attractive alternative to strictly CMOS to design reliable rad-hard and cold-capable circuits. In this dissertation proposal, the design of an SRAM bit cell with ECL peripherals will be explored using a BiCMOS design approach to achieve radiation-hardened and cold-capable circuits.

%{--Insert figure for general topology--}

%There are many different variations to the Flash ADC which provide the designer an advantage over other configurations at the cost of one or more design metrics.  Voltage folding and time folding are common examples.  Folding techniques allow the designer to partition the signal such the each partition can be evaluated as a sub-ADC, which greatly reduces the required resistor and comparator count.

%For this project, a general 4-bit Flash ADC was chosen, with a Ringamp-based comparator and bubble error correction logic implemented into the Thermometer to Binary conversion logic.  The Ringamp-based comparator is a low power option as the Ringamp is essentially a chain of inverters.  {Omar -  discuss bubble error correction and it's primary advantage.} 

%The remainder of this report is organized as follows: Section~\ref{section:Theory} will discuss the theory required for designing this 4-bit Flash ADC.  Section~\ref{section:Simulations} will discuss the simulation configuration and results.  And Section~\ref{section:Conclusions} will discuss the conclusions of this report along with design considerations for improvements to this ADC.

\section{Device Reliability in Extreme Temperatures}\label{section:Theory}
\subsection{Reliability of Planar Bulk MOSFET in Cryogenic Temperatures}\label{section_Theory}
Designing front-end application-specific integrated circuits (ASICs) for cryogenic temperature (approx {-$180^{\circ}$}{C}) is difficult if the requirements set for those circuits need to be low-noise, power efficient, can perform accurate signal processing, and have a lifetime of about 20 to 30 years \cite{YChen, Brookhavenppt}. The temperature models developed by most of the CMOS foundries for military-grade electronics usually have a maximum temperature range of {-$55^{\circ}$}{C} to {$125^{\circ}$}{C} and have the lifetime of the devices is about 10 years \cite{Brookhavenppt, Braga}. Although the most common failure mechanisms in MOSFET devices are thermal cycling, electromigration, time‐dependent dielectric breakdown, negative-bias temperature instability, and stress migration, have large temperature dependence but at cryogenic temperature, those effects are not very significant \cite{Brookhavenppt, Shaorui}. Hot carrier effects (HCE) is the sole mechanism left that can negatively impact the lifetime (10\% reduction in $g_m$) of the devices at cryogenic temperature of the CMOS devices by degradation or aging, mainly in the NMOS devices because usually NMOS has a lifetime two orders of magnitude shorter than PMOS, although the same mechanism is present in PMOS \cite{YChen, Brookhavenppt}. Through impact ionization, HCE at cryogenic temperature produces oxide trap charges and interface states in the gate oxide and at the $Si$-$SiO_2$ interfaces, respectively  \cite{Brookhavenppt, Shaorui, Mathesis}. In the 1980s, the constant voltage scaling in the CMOS devices exacerbated the hot-carrier effect (HCE) because of the large electric field. However, in the mid-1990s, the scaling of power supply voltage was reduced to save power and minimize reliability issues. This trend was slowed down due to the non-scalability of the sub-threshold slope at that time. But with the introduction of modern nodes, it is no longer an issue, i.e. supply voltages have continued to scale.  At cryogenic temperatures, the degradation or aging due to HCE is still a modern issue \cite{AnalogICReliability}.

This degradation or aging mechanism is not going to cause a sudden device failure in the circuit but the devices will fail gradually if the parameters like transconductance,  drain current, low-field-mobility, sub-threshold swing, threshold voltage, low-frequency noise, etc. get out of the specification range. Experiments by Shaorui Li et al \cite{Shaorui, Yuanke, Mathesis} showed that with careful electronic circuit design, this gradual degradation can be controlled both at room and cryogenic temperatures. It has been observed that the lifetime at 77K is the same as 300K if the $V_{DS}$ is reduced by 10\% at the 180 nm node \cite{Shaorui}. The device models that can capture noise performance, the static and dynamic response, and the CMOS devices and circuits lifetime accurately in cryogenic temperature ({-$180^{\circ}$}{C}) are important for designing HCE-protected front-end ASICs. To have a low power requirement on analog circuits, the models must also include the moderate inversion region, alongside strong and weak inversion \cite{Brookhavenppt}.

%\afterpage{
\begin{figure} 
         \centering 
         \includegraphics[scale=0.55]{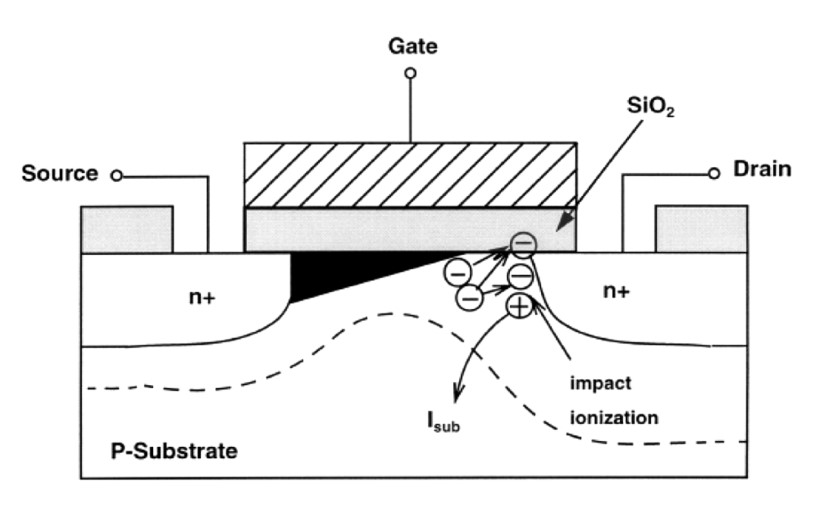}
         \caption[Schematic illustration of the stress of the Hot Carrier Effect (impact ionization) in an NMOS channel area. The substrate current is comprised of holes, created in the channel area by impact ionization. ]{Schematic illustration of the stress of the Hot Carrier Effect (impact ionization) in an NMOS channel area. The substrate current is comprised of holes, created in the channel area by impact ionization  \cite{Shaorui}.}
         \label{fig:nMOS_HCE}
\end{figure}
%\clearpage

If the electron acquires energy E $\gg$ kT (where, k=Boltzmann constant, and T=Temperature) in an NMOS device because of HCE, this can generate an electron-hole pair, $\phi_{i}$ $\approx$ 1.3eV \cite{Tam}, which can start an impact ionization process as shown in  Fig.~\ref{fig:nMOS_HCE}. As $V_{ds}$ creates a lateral electric field, which pushes the electrons towards the drain, and the holes towards the substrate \cite{Shaorui}. Then, the substrate current is given by \cite{Shaorui},
\begin{equation}
%\large
\label{eq:I_sub}
    I_{sub} = C_{1} I_{ds} exp(\frac{-\phi_{i}}{q \lambda  E_{m}})
\end{equation}

where $C_{1}$ = constant, $\phi_{i}$= the critical energy to generate an electron-hole pair
through impact ionization, $I_{ds}$ = the drain-source current, $q$ = the electron charge, $\lambda$ = the electron mean free path, and $E_{m}$ = the maximum channel lateral electric field proportional to the difference between $V_{ds}$ and $V_{dsat}$ i.e. $E_{m}$ $\propto$ ($V_{ds}$-$V_{dsat}$). Among these high-energy (i.e. E $\gg$ kT) electrons, a tiny fraction can inject them-self into $SiO_2$ via hot carrier injection and create an interface state (e.g. an acceptor-like trap) \cite{Shaorui} at the $Si$-$SiO_2$ interface, where for electron $\phi_{it}$ $\geq$ 3.7 eV (for hole $\phi_{it}$ $\geq$ 4.6 eV) as shown in Fig.~\ref{fig:nMOS_HCE}. HCE is strongly affected by these interface states and causes degradation in transistor characteristics (i.e. transconductance, mobility, threshold voltage, subthreshold swing, and intrinsic gain) \cite{Shaorui, Mathesis}.  The generation of interface states by HCE is given by \cite{Mathesis},
\begin{equation}
%\large
\label{eq:DeltaN_it}
    \Delta N_{it} = C_{2}(t \frac{I_{ds}}{W} exp(\frac{-\phi_{i}}{q \lambda  E_{m}}))^n
\end{equation}

where $C_{2}$=constant, $\Delta$$N_{it}$ = the interface states generated by HCE, $t$ = the hot carrier stress time, $\frac{I_{ds}}{W}$ = the drain current density, $\phi_{it}$ = the minimum energy to create interface state, W = the channel width, and n = constant. Equation \ref{eq:DeltaN_it} shows that $\Delta$$N_{it}$ grows over time. The device lifetime is the time needed to deteriorate any important parameter like transconductance reduced by 10\%. This can be deduced from Equation \ref{eq:DeltaN_it} \cite{Mathesis},

\begin{equation}
%\large
\label{eq:lifetime}
    \tau = C_{3} \frac{W}{I_{ds}} exp(\frac{\phi_{i}}{q \lambda  E_{m}})
\end{equation}
where $C_{3}$ is a constant.

In equation \ref{eq:lifetime}, the lifetime of a MOSFET is predicted to be much shorter at cryogenic temperature because the mean free path $\lambda$ gets longer as the phonon scattering reduces with decreasing temperature \cite{Chen}.

It has been established in the literature \cite{Chen, Hu}, that monitoring the substrate current for all HCEs is the best process to estimate a device lifetime as both (by electrical and optical) kinds of HCEs are run by the same driving force, i.e. the largest electric field $E_{m}$ in the channel effective mostly at the channel's drain side as shown in Fig.~\ref{fig:nMOS_HCE}. The lifetime, $\tau$ can be deduced from the substrate's current from the equation \ref{eq:I_sub} and \ref{eq:lifetime} by substituting $q$$\lambda$$E_{m}$,

\begin{equation}
%\large
\label{eq:lifetime_1}
    \tau = H \frac{1}{I_{ds}/W} (\frac{I_{sub}}{I_{ds}})^{{-\phi_{it}}/{\phi_{i}}}
\end{equation}

Here H (in As/µm) = a constant that depends on parameters like the temperature, the channel length, the device fabrication method (drain doping, interface quality, etc.), and the metrics used to define lifetime \cite{Mathesis}. Now, from case to case, the proportionality constant H can vary, the relationship in equation \ref{eq:lifetime_1} can be expressed in terms of lifetime as,

\begin{equation}
%\large
\label{eq:lifetime_2}
    \frac{\tau I_{ds}}{W} \propto \frac{1}{(\frac{I_{sub}}{I_{ds}})^\alpha}
\end{equation}

Here, the exponent $\alpha$ = {$\phi_{it}$}/{$\phi_{i}$} is the ratio of the critical energy of the electron to produce an interface state, $\phi_{it}$ ($\approx$ 3.7-4.1eV), and the critical energy to generate an electron-hole pair through impact ionization, $\phi_{it}$ ($\approx$ 1.3 eV ), has a value between 2.9 $m$ 3.2. The lifetime of the device depends on the substrate current that comes from these two critical energy ratios, which is not temperature-dependent \cite{Mathesis}.

\begin{figure} 
         \centering 
         \includegraphics[scale=0.60]{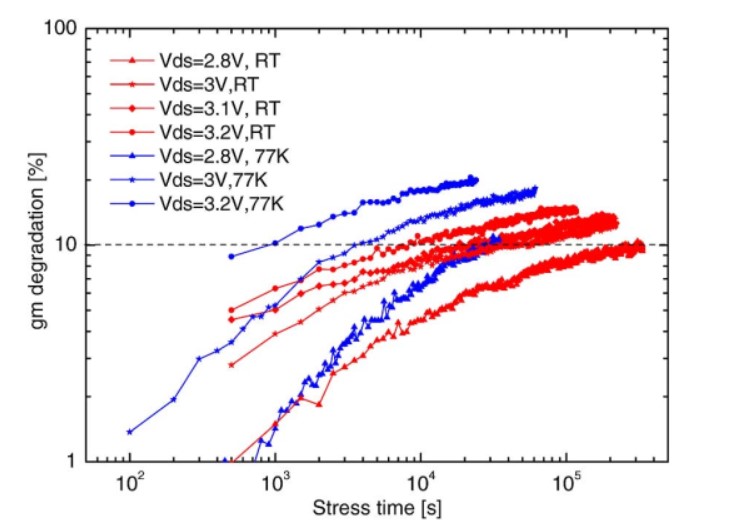}
         \caption{Under accelerated stress, the transconductance degradation vs. time of an NMOS transistor (W=5$\times$2 $\mu$m, L= 180 nm) is measured at 300K and 77 K \cite{Shaorui}.}
         \label{fig:NMOS_trnsconductance}
\end{figure}

\begin{figure} 
         \centering 
         \includegraphics[scale=0.60]{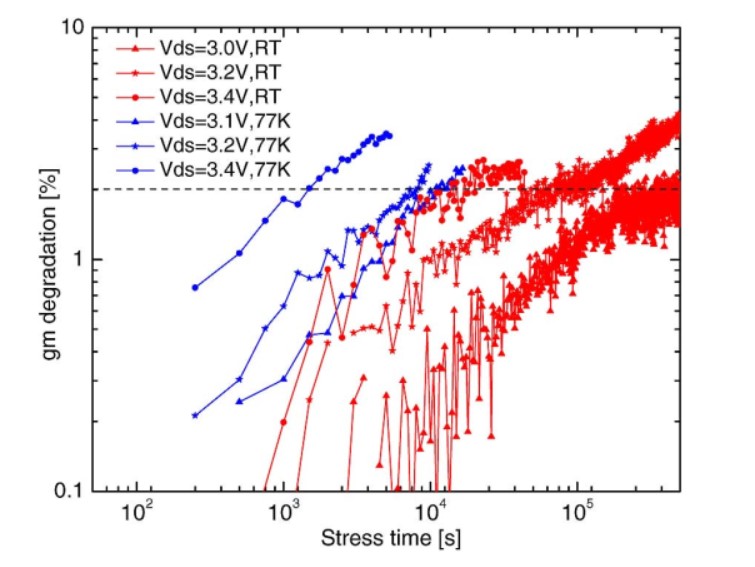}
         \caption{Under accelerated stress, the transconductance degradation vs. time of a PMOS transistor (W=5$\times$2 $\mu$m, L= 180 nm) is measured at 300K and 77 K \cite{Shaorui}.}
         \label{fig:PMOS_trnsconductance}
\end{figure}

 Y. Chen and et. al. \cite{YChen} had done accelerated lifetime tests with a 500 nm node and  J.R. Hoff \cite{Hoff} with a 130 nm node. Both observed similar degradation of lifetime in NMOS at cryogenic temperature. Y. Chen and et. al. \cite{YChen} reported the PMOS lifetime has 2 orders of magnitude larger than the NMOS lifetime under hot carrier aging. But Y. Chen and et. al. \cite{YChen} and  J.R. Hoff \cite{Hoff} focused their study on NMOS transistors and did not report the PMOS data separately for direct comparison.  Shaorui Li et al \cite{Shaorui, Mathesis} have done accelerated lifetime tests, which are well accepted throughout the foundries. They tested devices from the TSMC 180-nm process at 300 K and 77 K. Here, the transistor is tested by putting it under large electric field stress with a large $V_{DS}$, then the hot carrier electron degradation will reduce the lifetime within an acceptable observable timeframe, by setting the $V_{DS}$ much larger than the nominal voltage (1.8V breakdown limit for $L_{min}$ = 180 nm). In Shaorui Li et al \cite{Shaorui, Mathesis} experiments, the degradation or aging in PMOS is much slower than in NMOS devices under accelerated lifetime test. Fig.~\ref{fig:PMOS_trnsconductance} shows at 300 K and 77 K, at the voltage from 3 to 3.4 V, the PMOS devices are going under stress, whereas, in Fig.~\ref{fig:NMOS_trnsconductance} NMOS devices are at lower range of 2.8 to 3.2 V stress voltage, but PMOS only having a $m$ 2$\%$ degradation at stress time, in comparison, NMOS is experiencing a $m$ 10$\%$ degradation. PMOS is in the region of early-mode degradation and has degradation much slower than NMOS \cite{Shaorui}, which was anticipated by Y. Chen et al in \cite{YChen}. 
%\afterpage{
%\clearpage}

%\afterpage{
\begin{figure} 
         \centering 
         \includegraphics[scale=0.65]{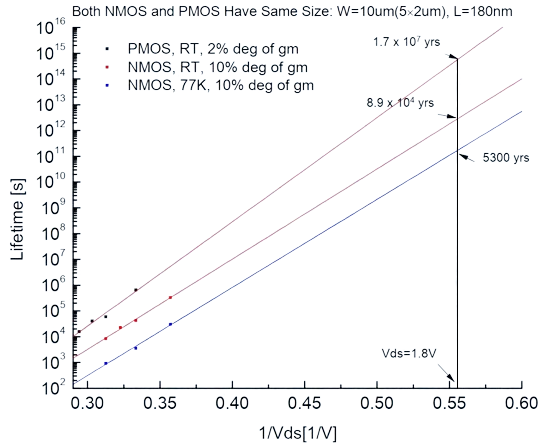}
         \caption{The stress measurement results are used to project 
         lifetime \cite{Brookhavenppt}.}
         \label{fig:Lifetime_vs_(1/V_{ds}}
\end{figure}
%\clearpage}

After measuring the transconductance degradation in Fig.~\ref{fig:NMOS_trnsconductance} and Fig.~\ref{fig:PMOS_trnsconductance}, a stress plot can be produced from the measured data at $m$ 10$\%$ for NMOS and $m$ 2$\%$ for PMOS to calculate the characteristic slope of the generation of the interface state $\alpha$ in Equation \ref{eq:lifetime_2}. In the experiment, Shaorui Li et al. measured slopes at both 300 K and 77 K, which are found to be very close to their theoretical counterpart, i.e., $\alpha$ $\approx$ 3.
The constant H in Equation \ref{eq:lifetime_1} is a function of $I_{ds}$ and $V_{ds}$ at low voltage but in these measurements the characteristic slope remains unaffected because
$I_{ds}$ and $V_{ds}$ only varied in a limited range ($<$ 5\%) \cite{Shaorui, Mathesis}.
From Equation $\ref{eq:lifetime}$, that is, $\ln{(\frac{\tau I_{ds}}{W})}$ $\propto$ $\frac{\phi_{i}}{(q \lambda  E_{m})}$ $\propto$ {1}/{($V_{ds}$-$V_{dsat}$)} can estimate the lifetime of the stress points at the core voltage. The pinch-off voltage $V_{dsat}$ can be ignored at low power with moderate and weak inversion for analog front-end devices. Then, the lifetime vs. 1/$V_{ds}$ is estimated as shown in Fig.~\ref{fig:Lifetime_vs_(1/V_{ds}}. The projected lifetime for NMOS is ~2 orders of magnitude shorter than that of PMOS \cite{Brookhavenppt, Shaorui}.

\subsection{SiGe HBTs at Cryogenic Temperature}

Bandgap engineering with epitaxially grown Ge in the Si substrate yields the Silicon Germanium (SiGe) heterojunction bipolar transistor (HBT), which is a very attractive alternative to the Si bipolar junction transistor (BJT). Due to its 100\% compatibility with Si manufacturing, it retains excellent analog and RF characteristics, that is, higher gain, better matching, lower broadband noise, higher transconductance per unit area,  lower 1/f noise, higher output resistance, higher $f_{T}$ and $f_{max}$, etc. at a very harsh environment such as high radiation and extreme cold \cite{COLDTech}. Examples of such harsh environments are very cold temperature operations down to 77 K or 4.2 K or below, very high temperature operations up to 573 K ($+300^{\circ}${C}), very wide and/or cyclic temperature swings like at lunar surface temperature swings $+120^{\circ}${C} to $-230^{\circ}${C}  from day to night, operation in high radiation (on Europa surface 5 Mrad), or combination of few or all of the above, which will be the worst case scenario \cite{Cressler_Extreme}. SiGe HBTs, and the circuits designed with it have the potential to tolerate all the extreme environmental events mentioned above with some modification in the fabrication process of Si BJT process, while providing considerable advantages to implement commercial and defense applications in the integrated circuit and system level \cite{Cressler_Extreme}.

%\afterpage{
\begin{figure} 
         \centering 
         \includegraphics[scale=0.50]{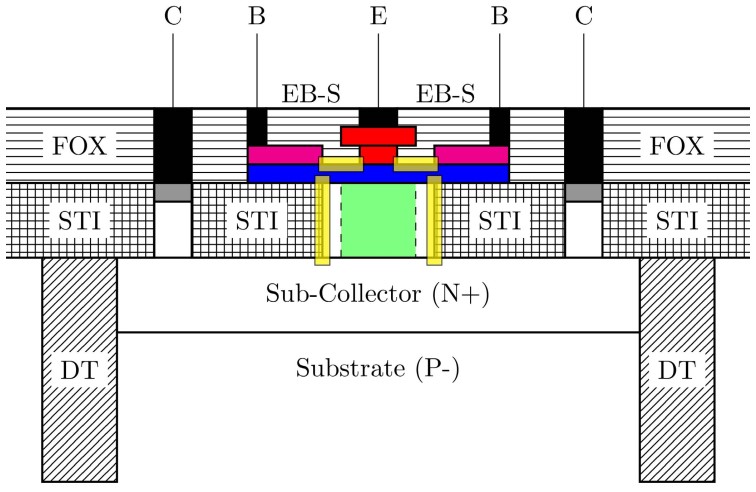}
         \caption[ SiGe HBT cross-section. Where, EB-S = emitter-base spacer, FOX = field oxide, DT = deep trench, STI = shallow-trench isolation, TID-sensitive regions are pinpointed in yellow. Intrinsic base (\(P^{+}\)) is colored in blue , polysilicon emitter (\(N^{++}\)) is colored in red, selectively-implanted collector (\(N^{-}\)) is colored in green, extrinsic base (\(P^{++}\)) is colored in magenta]{SiGe HBT cross-section. Where, EB-S = emitter-base spacer, FOX = field oxide, DT = deep trench, STI = shallow-trench isolation, TID-sensitive regions are pinpointed in yellow. Intrinsic base (\(P^{+}\)) is colored in blue , polysilicon emitter (\(N^{++}\)) is colored in red, selectively-implanted collector (\(N^{-}\)) is colored in green, extrinsic base (\(P^{++}\)) is colored in magenta \cite{Jeff}.}
         \label{fig:SiGe_HBT_Cross_section}
\end{figure}

\begin{figure} 
         \centering 
         \includegraphics[scale=0.40]{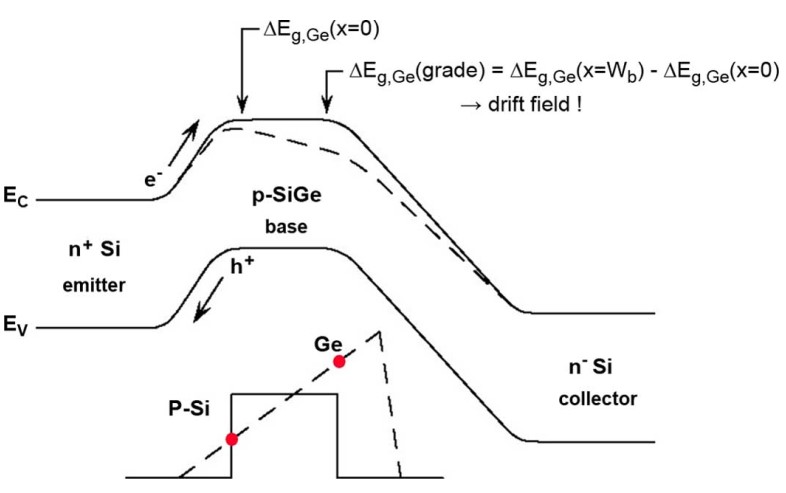}
         \caption[Schematic comparison in the energy band diagrams of a Si BJT and a graded-base SiGe HBT, where both are doped identically and have forward-active mode biasing.]{Schematic comparison in the energy band diagrams of a Si BJT and a graded-base SiGe HBT, where both are doped identically and have forward-active mode biasing \cite{Cressler2013}.}
         \label{fig:Band_diagram_drift_field_SiGe_HBT}
\end{figure}
%\clearpage}

 Fig.~\ref{fig:SiGe_HBT_Cross_section} shows the schematic of a representative cross-section of a SiGe Heterojunction Bipolar Transistor (HBT) with emitter areas ($A_E$) of  0.1 $\times$ 4  ${\mu m}^2$, with two collectors and two base contacts.  They have a current gain ($\beta$) of 500 at 300 K but at 77 K, $\beta$ goes up close to 1000.  The unity gain cutoff frequency ($f_T$) is 300 GHz (at 90 nm) at 300 K, at cold (77 K) $f_T$ it increases to 450 GHz \cite{Jeff}. The device in Fig.~\ref{fig:SiGe_HBT_Cross_section} is adapted from the GlobalFoundries 90 nm SiGe BiCMOS technology platform \cite{Jeff}. SiGe HBTs usually have extremely fast device speeds, depending on the scaling path moving forward to ensure the evolution of electronic applications. In extreme environments like high radiation and cryogenic temperatures, the vertical profile of the device has a very strong effect on the performance as well as the technology node. 

Fig.~\ref{fig:Band_diagram_drift_field_SiGe_HBT} shows the contrast in the energy band diagrams of a Si BJT and a SiGe HBT, where both are doped identically and in forward-active mode. To simplify the analysis, let's consider the emitter, base, and collector region of an ideal, graded-base SiGe HBT is doped at a constant level. In this device, Ge doping is added linearly, 0$\%$ at the metallurgical emitter-base (EB) junction to an arbitrary pick value at near the metallurgical collector-base (CB) junction, and from there swiftly goes back to 0$\%$ Ge. This doping profile shifts up when heavy doping is applied to both the emitter and collector. Fig.~\ref{fig:Band_diagram_drift_field_SiGe_HBT} shows the trimming of base bandgap by the graded Ge at the EB edge of the base ($\Delta$$E_{g,Ge}$(x=0)), and the CB edge of the base  ($\Delta$$E_{g,Ge}$(x=$W_b$)). A quasi-drift field ($\Delta$$E_{g,Ge}$(x=$W_b$))-($\Delta$$E_{g,Ge}$(x=0)/$W_b$) is created by this graded Ge in the base section \cite{SiGeHBT2003}.
%This graded Ge throughout the base creates a built-in quasi-drift field ($\Delta$$E_{g,Ge}$(x=$W_b$))-($\Delta$$E_{g,Ge}$(x=0)/$W_b$)  \cite{SiGeHBT2003}.

The DC characteristics change in the SiGe HBT induced by Ge-based band edge changes can be visualized with forward biased emitter-base junction, the inserted electrons disperse through the base and move into the collector-base junction's electric field, contributing to collector current. On the other hand, because of the forward bias of the emitter-base junction, the holes are getting back-injected from the base to the emitter. As the emitter doping is heavier than the base, the forward-injected electron density is much larger than the back-injected holes, creating a current gain \cite{ Cressler2013}. 

%\afterpage{
\begin{figure} 
         \centering 
         \includegraphics[scale=0.50]{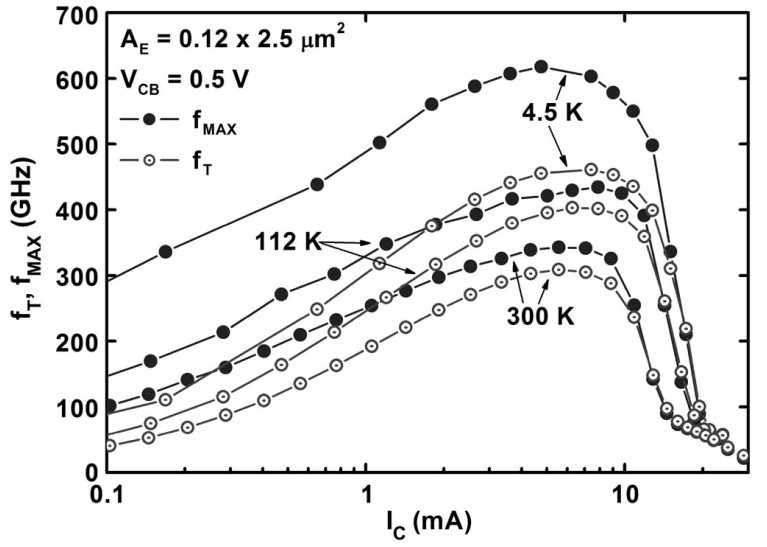}
         \caption[ Cutoff frequencies $f_{T}$ and maximum oscillation $f_{max}$ vs. collector bias current measured at temperature 300 K, 112 K, 4.5 K of a fourth-generation SiGe HBT prototype.]{Cutoff frequencies $f_{T}$ and maximum oscillation $f_{max}$ vs. collector bias current measured at temperature 300 K, 112 K, 4.5 K of a fourth-generation SiGe HBT prototype \cite{Jiahui}.}
         \label{fig:Max_oscillation}
\end{figure}

\begin{table}[]
\centering 
\caption{At 300 K, 112 K, and 4.5 K, the summary of the different parameters of a fourth-generation SiGe HBT prototype \cite{Jiahui}.}
\begin{tabular}{|cccc|}
\hline
%\multicolumn{1}{|c|}{Parameters}      & \multicolumn{1}{c|}{\begin{tabular}[c]{@{}c@{}}\end{tabular}} & \multicolumn{1}{c|}{\begin{tabular}[c]{@{}c@{}}\end{tabular}} & \begin{tabular}[c]{@{}c@{}}\end{tabular} \\ \hline
\multicolumn{1}{|c|}{Temperature (K)}                & \multicolumn{1}{c|}{300}                                                 & \multicolumn{1}{c|}{112}                                                      & 4.5                                                 \\ \hline
\multicolumn{1}{|c|}{Peak {$\beta$} }               & \multicolumn{1}{c|}{827}                                                  & \multicolumn{1}{c|}{6504}                                                      & 7693                                                  \\ \hline
\multicolumn{1}{|c|}{Peak  $g_{m}$  (mS)} & \multicolumn{1}{c|}{72}                                                 & \multicolumn{1}{c|}{110}                                                    & 113                                                \\ \hline
\multicolumn{1}{|c|}{Peak $f_{max}$  (GHz)}     & \multicolumn{1}{c|}{343}                                         & \multicolumn{1}{c|}{434}                                            & 618                                         \\ \hline
\multicolumn{1}{|c|}{Peak $f_{T}$ (GHz)}         & \multicolumn{1}{c|}{309}                                         & \multicolumn{1}{c|}{403}                                             & 463                                            \\ \hline
\multicolumn{1}{|c|}{Transit Time $\tau_F$ (fs)}         & \multicolumn{1}{c|}{420}                                         & \multicolumn{1}{c|}{330}                                            & 300                                           \\ \hline
\multicolumn{1}{|c|}{$V_{BE}$  at  $f_{max}$  (V)}  & \multicolumn{1}{c|}{0.90}                                           & \multicolumn{1}{c|}{1.04}                                              & 1.06                                           \\ \hline
\multicolumn{1}{|c|}{$I_{C}$ at peak $f_{max}$ (mA)}  & \multicolumn{1}{c|}{5.6}                                           & \multicolumn{1}{c|}{7.9}                                              & 4.8                                          \\ \hline
\multicolumn{1}{|c|}{$BV_{CEO}$ (V)}  & \multicolumn{1}{c|}{1.70}                                           & \multicolumn{1}{c|}{1.63}                                              & 1.62                                          \\ \hline
\multicolumn{1}{|c|}{$BV_{CBO}$ (V)}  & \multicolumn{1}{c|}{5.6}                                           & \multicolumn{1}{c|}{5.6}                                              & 5.6                                          \\ \hline
\end{tabular}
%\caption{At 300 K, 112 K, and 4.5 K, the summary of the different parameters of a fourth-generation SiGe HBT prototype \cite{Jiahui}.}
\label{table:Parmeter_HBT_Temp}
\end{table}
%\clearpage}

Because the graded Ge-induced drift field throughout the base is in the opposite direction as the emitter to the collector it helps to speed up the minority electrons injection from the base to the collector, which helps to improve the AC characteristics of the SiGe HBT. The electron transport gets a boost because the added drift field is significantly large enough to speed up the diffusive transport in minority electrons and reduce the base transit time. If a 50-nm neutral base 
has linear Ge grading with a modest 10$\%$ Ge content at the peak will produce an electric field of 75 mV/50 nm =15 KV/cm, which will produce enough acceleration for the electrons to take them close to the saturation velocity ($v_s$ $\approx$ 1-2 $\times$ ${10}^7$ cm/sec). As in the Si BJT, the base transit time restricts the frequency response, the addition of Ge-induced drift field in SiGe HBTs will contribute significantly to the improvement of the frequency response. At the EB junction, the band offset by graded Ge will increase the collector current exponentially if compared to SI BJT. The added emitter charge storage delay time also helps the frequency response of the SiGe HBT \cite{ Cressler2013}. For AC response, the $f_{T}$ is given by \cite{SiGeHBT2003},

\begin{equation}
 %\large 
\label{eq:f_T}
    f_{T} = \frac{1}{2\pi} [\frac{kT}{q I_C}(C_{te}+ C_{tc})+\tau_b+ \tau_e+ \frac{W_{CB}}{2v_{sat}}+r_cC_{tc}]^{-1} 
\end{equation}

Where, the intrinsic transconductance $g_m$ ={q$I_C$}/{kT} at low-injection ($g_m$= {$\partial$ $I_C$}/{$\partial$ $V_{BE}$}),  $C_{te}$ and $C_{tc}$ = EB and CB deplation capacitances, $\tau_b$ = the base transit time, $\tau_e$ = the emitter charge storage delay time, $v_{sat}$ = the saturation velocity, and $r_c$ = the dynamic collector resistance. 

Fig.~\ref{fig:Max_oscillation} and Table ~\ref{table:Parmeter_HBT_Temp} show at cold temperatures most of the positive effects get amplified because of the bandgap engineering in SiGe HBTs. So, the cryogenic operation can be used as the ultimate scaling limit, and at 4.5 K, with a breakdown voltage above 1.3 V, the $f_{max}$ has a record measured value of 600 GHz and $f_{T}$+ $f_{max}$ has a combined value above 1 THz. At 300 K, SiGe HBTs can have a peak $f_{T}$ and $f_{max}$ above 500 GHz \cite{Cressler_Extreme}. 
As a minority carrier device, the Ge-induced engineered band edge change in the SiGe HBTs will generate thermally activated currents with exponential dependence on E/kT.
Then, from the fundamental physics of the SiGe HBT, it is known that cold temperature positively improves all device metrics from a circuit point of view, like transconductance, $f_{T}$, $f_{max}$, current gain, broadband noise, etc \cite{SiGeHBT2003, SiGeHetro2006}. This is great news for any semiconductor devices 
designed for wide-temperature operation. This opens up the options to solve the issues with the cooling of homojunction bipolar transistors (BJT) causing carrier diffusivity degradation,  base freezeout, and bandgap-narrowing-induced gain degradation, whereas cooling the SiGe HBT improves its characteristics. Usually, if the SiGe HBTs are fabricated accurately at room temperature (300 K), they will perform better in cold temperatures \cite{Cressler_Extreme}. This claim of the superior performance of SiGe HBTs in cold temperature is supported with sufficient data in the literature \cite{CresslerSiGe77k1, CresslerSiGe77k2, CresslerSiGe77k3, CresslerSiGe77k4, CresslerSiGe77k5, CresslerSiGe77k6, CresslerSiGe77k7, CresslerSiGe77k8}.  

%\afterpage{
\begin{figure} 
         \centering 
         \includegraphics[scale=0.50]{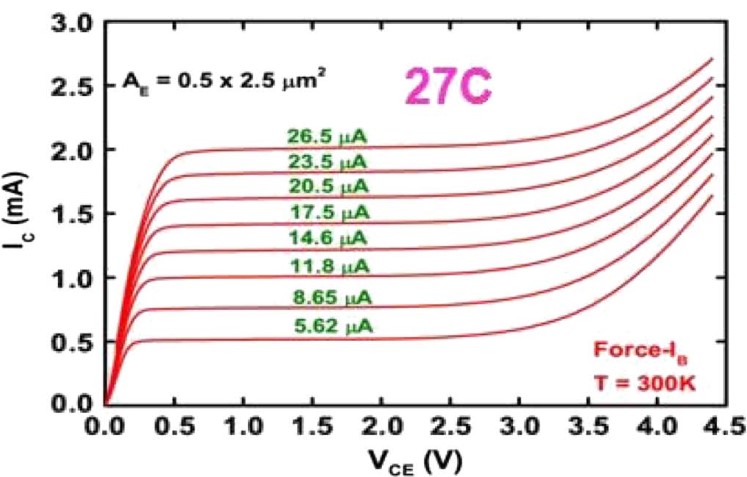}
         %\label{fig:Output_Characteristics_SiGe_HBT_27C}
\end{figure}

\begin{figure} 
         \centering 
         \includegraphics[scale=0.50]{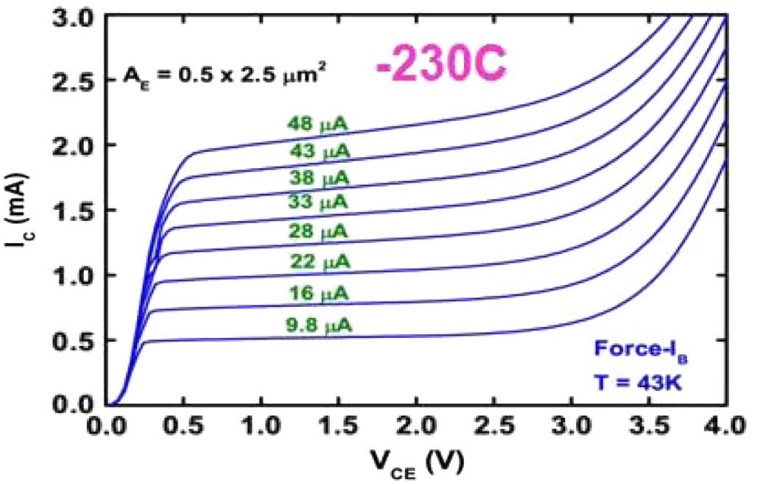}
         \caption[At temperature {$27^{\circ}$}{C} and {-$230^{\circ}$}{C}, a first-generation SiGe HBT is exhibiting temperature invariance in its output characteristics.]{At temperature {$27^{\circ}$}{C} and {-$230^{\circ}$}{C}, a first-generation SiGe HBT is exhibiting temperature invariance in its output characteristics \cite{Cressler_Extreme}.}
         \label{fig:Output_Characteristics_SiGe_HBT_27C_-230C}
\end{figure}
%\clearpage} 

Fig.~\ref{fig:Output_Characteristics_SiGe_HBT_27C_-230C} is the plot of the measured output characteristics of a commercial first-gen SiGe HBT at {$27^{\circ}$}{C} (300 K) and {$-230^{\circ}$}{C} (43 K).  The lowest temperature found on the Earth's moon is {$-230^{\circ}$}{C} (43 K). Both pictures in 
Fig.~\ref{fig:Output_Characteristics_SiGe_HBT_27C_-230C} are the measured output characteristics {$250^{\circ}$}{C} apart but looks more or less same current gain, output conductance, current drive, and breakdown voltage at the cold temperature ({$-230^{\circ}$}{C}). This implies that SiGe HBTs are at least a first-order “temperature-invariant” device platform. These temperature-invariant characteristics over wide temperatures are very impactful on circuit design techniques and distinctive among semiconductor devices. From the experimental verification, it has been observed that if SiGe HBTs are fabricated well, then they will demonstrate the same temperature invariance in all performance metrics at the circuit level \cite{Cressler_Extreme}. This is the opposite of conventional CMOS because it suffers HCE at cold temperatures. SiGe HBTs also have some second-order effects like barrier effects,  higher leakage, and negative differential resistance effects, but if the devices are well fabricated then they remain second-order \cite{CresslerSiGe10}. 

\begin{figure} 
         \centering 
         \includegraphics[scale=0.450]{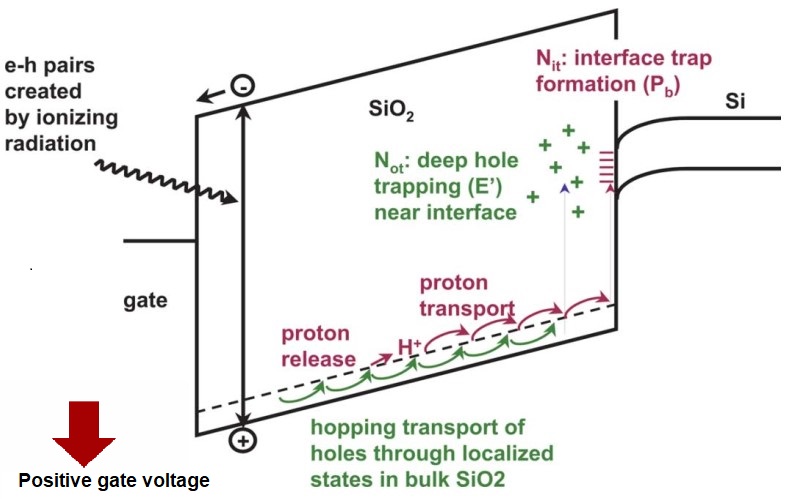}
         \caption[Radiation-induced charge generation processes are explained in the band diagram of a positively gate biased MOS cap.]{Radiation-induced charge generation processes are explained in the band diagram of a positively gate biased MOS cap \cite{Moscap,TIDMOSFETTI}.}
         \label{fig:MOScap_Banddiagram_under_redaition}
\end{figure}

\begin{figure} 
         \centering 
         \includegraphics[scale=0.350]{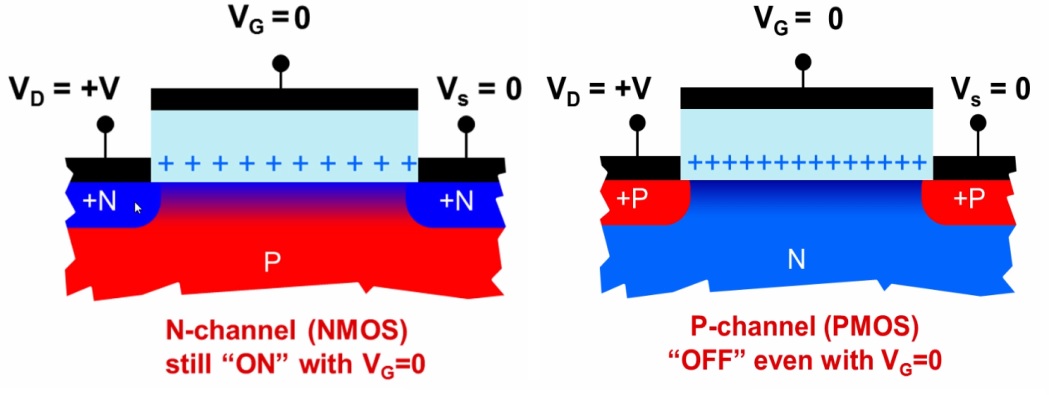}
        \caption[Domination of the positively charged traps in most of the MOSFET 
        behavior due to TID.]{Domination of the positively charged traps in most 
        of the MOSFET behavior due to TID \cite{TIDMOSFETTI}.}
        \label{fig:MOSFET_under_TID}
\end{figure}

\section{MOSFET and SiGe HBTs under High Radiation}

\label{sec:High_Radiation}

The goal of the National Aeronautics and Space Administration's (NASA) ``Concepts for Ocean Worlds Life Detection Technology (COLDTech)" Program is to start the development of environmentally invariant electronics for the exploration of the moons and other planets with liquid water oceans in our solar system to check future habitability and collect samples as the biosignatures of extraterrestrial life \cite{Nasaow}. Among the world with liquid water oceans under its thick ice shelf, Europa is of particular importance because it has the smallest distance from the surface to the ocean beneath (10 km), and the proposed Europa lander mission is planning to do scientific measurements from the surface of the Europa \cite{Jeff, Europalander1}. The magnetosphere around Jupiter traps electrons causing a significant electron fluence, with a total ionizing dose (TID) for any electronic system traveling around in a spacecraft \cite{Jeff}. Usually, the electronics systems have to face a dose of few Mrad(Si) even though the dose is exhausted by Europa's magnetic field, atmosphere,
and the shielding inside the spacecraft \cite{Jeff, Geophysics07}. So, any future missions on Europa's surface are going to face severe radiation and need to implement surface electronics that can operate within the mission parameters during the exposure. 

\subsection{Effect of Total Ionizing Dose (TID) in MOSFET}

 In space, electronic devices with different kinds of oxides and insulators can have a considerable buildup of charges due to ionizing radiation, leading to device degradation and failure over time. Usually, the two main sources of these radiation-induced total ionizing dose (TID) effects are electrons and protons, which generate charges. These charges create oxide-trap and interface-trap inside the oxide layer and at the $Si$-$SiO_2$ interface, respectively with large shifts in threshold voltages and increases in leakage currents \cite{Moscap}. The band diagram is shown in Fig.~\ref{fig:MOScap_Banddiagram_under_redaition}, when a p-substrate capacitor has a positive gate bias.

Based on the electric field magnitude and the incident radiation energy, some of these electron-hole pairs are subjected to recombination instantaneously. The rest of the electrons are drifted to the positively biased gate electrodes from the oxides within a few picoseconds,  but the holes approach the $Si$-$SiO_2$ interface through a hopping transport mechanism from a localized shallow trap site to another due to the disorder in the oxides. These electron-hole pairs that get away from recombination are referred to as the electron-hole yield or the charge yield. As the holes move towards the $Si$/$SiO_2$ interface, a portion of them get confined in deep hole traps like oxygen interstitials, oxygen vacancies, etc. inside the oxide layer and from the positive oxide-trap charges. The remaining holes react with the hydrogen, trapped in the oxygen defects, and produce hydrogen ions (protons).

As moving toward the $Si$/$SiO_2$ interfaces, the protons rupture the inert hydrogen dangling bonds and produce electrically active and amphoteric interface traps. These interface traps are usually positive in NMOS and negative in PMOS transistors. There are also other types of charge build-up due to radiation like inside the silicon-on-insulator (SOI) buried oxides, field oxide, alternate dielectrics, etc., which contribute to device degradation and circuit failure. When positive charges get trapped near the $Si$/$SiO_2$ interface, it can cause an inversion (or prevent any inversion) in the P-type (or N-type) Si substrate in the NMOS (or PMOS) transistors resulting in a large leakage current (or no ON current) to flow for NMOS (or PMOS), while $V_{GS}$= 0 V, i.e. the OFF state for NMOS or ON state for PMOS, as explained in Fig.~\ref{fig:MOSFET_under_TID}. As the technology nodes are scaling down, the gate oxides are getting thinner, and the radiation-induced IC degradation is becoming dominant due to charge build-up in the SOI-buried oxides and field oxides. All these contribute to a soaring static power supply current $I_{DD}$ and may lead to chip failure \cite{Moscap, Moscap1}.

\begin{figure} 
         \centering 
         \includegraphics[scale=0.50]{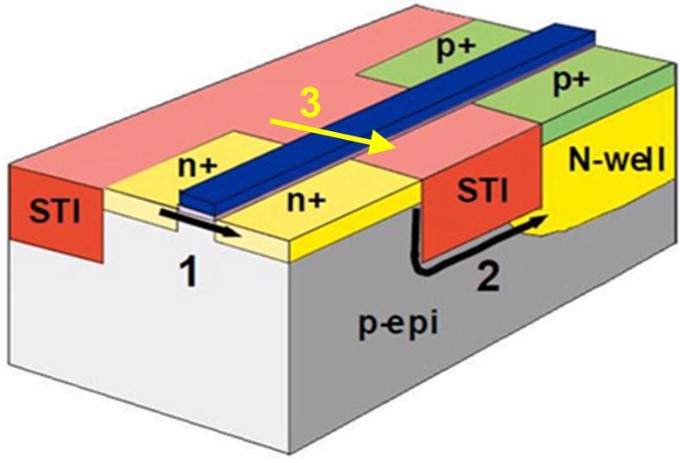}
         \caption{Possible leakage paths in a CMOS IC due to shallow trench isolation (STI) technology \cite{Moscap,TIDMOSFETTI}.}
         \label{fig:Leakage_Path_STI_1}
\end{figure}

\begin{figure} 
         \centering 
         \includegraphics[scale=0.450]{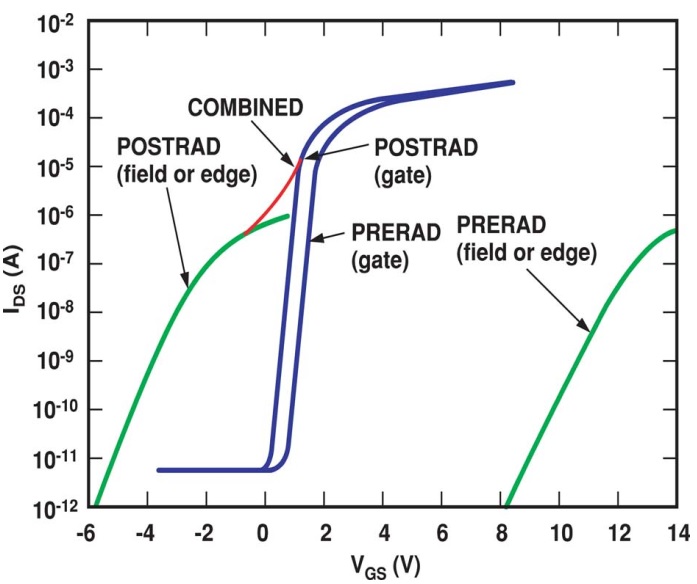}
        \caption{The gate oxide transistor has increased leakage current because there is a huge negative threshold-voltage shift in the parasitic field-oxide transistor after getting exposed to the radiation \cite{Moscap, TIDMOSFETTI}.}
        \label{fig:IV_CURVES_for_GateOxide_Transistor}
\end{figure}

If there is a large concentration of the interface trap charges near the $Si$/$SiO_2$ interface, that may cause reduced carriers mobility and increased NMOS transistor's threshold voltage, which contributes to the reduction of the transconductance ($g_m$) and deterioration of the timing parameters. TID will create a leaky NMOS, which will become harder to turn off (i.e. channel is still on, even though the gate is off) over time as the dose exposure continues to increase. For PMOS, there will be no leakage current due to TID because in the oxide layer, the presence of positive charges will effectively turn off PMOS but the ON current may be affected over time, i.e. it will be harder to turn on \cite{Moscap, TIDMOSFETTI}. 

Fig.~\ref{fig:Leakage_Path_STI_1} shows a CMOS circuit with complementary NMOS and PMOS devices combined with three feasible leakage paths because of shallow trench isolation (STI). Path 1 is the leakage path at the gate-oxide transistor edge that goes between the source and drain. Leakage path 2 occurs between the $n^+$ regions of an NMOS transistor (i.e. includes both source and drain) and the n-well of the neighboring PMOS transistor, i.e. a leakage path under the STI isolation to the n-well. Path 3 will be active, if the STI, the isolating layer or insulating layer between the two transistors (PMOS and NMOS) becomes charged. As STI has a very large volume, it can capture a large amount of charge, and then under an electric field, all the charges will accumulate at the edge of the channel.  As the radiation increases, these leakage currents also increase the DC supply current of an IC. However, the charge build-up inside the field oxides is mostly positive under high radiation, the leakage current mostly affects the NMOS transistors \cite{Moscap, TIDMOSFETTI}.  

Fig.~\ref{fig:IV_CURVES_for_GateOxide_Transistor} illustrates how a gate oxide transistor gets affected by the effects of the excess leakage current because the polysilicon gate, a part of the field oxide, the gate oxide transistor's source and drain creates the parasitic field-oxide transistor, which is responsible for the excess leakage current. Fig.~\ref{fig:IV_CURVES_for_GateOxide_Transistor} shows $I_{DS}$ and $I_{fieldleakage}$ current vs $V_{GS}$ of an NMOS with and without a field-oxide leakage. The threshold voltage of the parasitic field oxide transistor before radiation is usually very massive (above 8V) but a huge negative threshold-voltage shift happens in the parasitic field-oxide transistor because the radiation creates positive charges that build up inside the thick field oxide. With sufficient shifts in the parasitic field oxide transistor's threshold voltage, there will be a $I_{fieldleakage}$, i.e. "OFF" state leakage current of the parasitic field-oxide transistor added to the $I_{DS}$ of the gate oxide transistor \cite{Moscap, TIDMOSFETTI}.

\begin{figure} 
         \centering 
         \includegraphics[scale=0.4500]{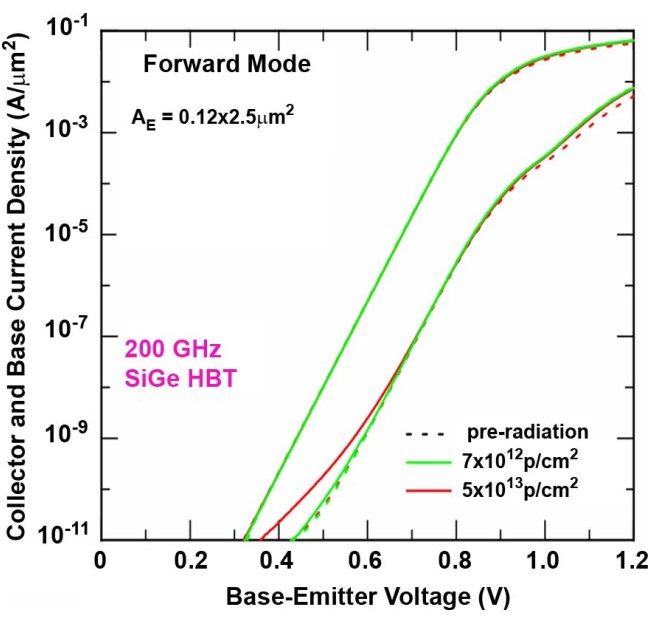}
         \caption[ Pre and post-radiation I-V characteristics of a 3rd-gen SiGe HBT, when subject to space-relevant 63-MeV protons with total dose up to multi-Mrad.]{Pre and post-radiation I-V characteristics of a 3rd-gen SiGe HBT, when subject to space-relevant 63-MeV protons with total dose up to multi-Mrad \cite{Cressler_Extreme}.}
         \label{fig:Current_voltage_characteristics_3rd_gen_SiGe_HBT}
\end{figure}

\begin{figure} 
         \centering 
         \includegraphics[scale=0.450]{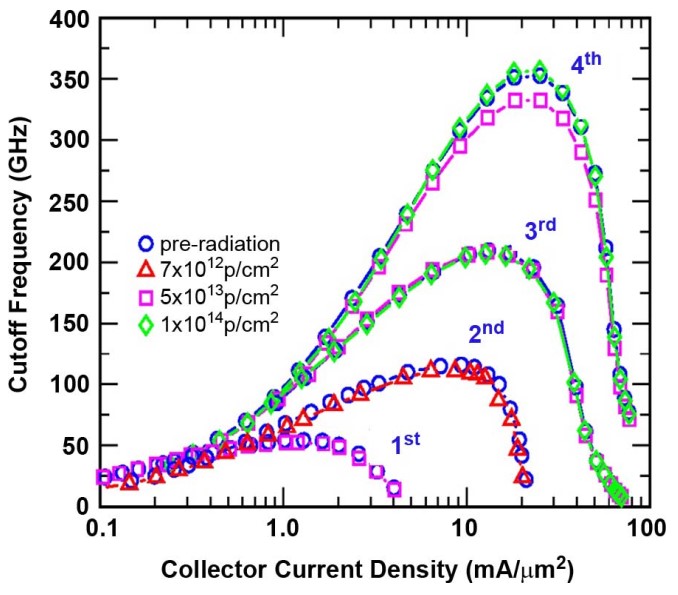}
        \caption[Pre and post-radiation cutoff frequency vs. bias current density for four different SiGe HBT generations, when subject to space-relevant 63-MeV protons with total dose up to multi-Mrad.]{Pre and post-radiation cutoff frequency vs. bias current density for four different SiGe HBT generations, when subject to space-relevant 63-MeV protons with total dose up to multi-Mrad \cite{Cressler_Extreme}.}
        \label{fig:Unity_gain_cutoff_frequency_4th_gen_SiGe_HBT}
\end{figure}

\subsection{Effect of Total Ionizing Dose (TID) in SiGe HBTs}

It has been established in the literature that  SiGe HBT can tolerate TID up to some 
multiMrad levels because of its internal built-in TID tolerance with dependence on the generation of the scaling node \cite{Cressler_Extreme, Babcock, Roldan, Zhang, Proton_tolerance, Sutton, Sutton1}. As Fig.~\ref{fig:Current_voltage_characteristics_3rd_gen_SiGe_HBT} shows, the I-V characteristics of a 3rd-generation SiGe HBT, post-exposure of a space-relevant 63-MeV proton with total dose up to multi-Mrad, the $I_{C}$ and the $I_{B}$ has very small shift from the pre-radiation time. A 63-MeV proton fluence of 5 $\times$ $10^{13}$ $p/cm^2$ will produce a dose of 6.7-Mrad, which is much more than SiGe HBTs will experience in Earth's orbit, even as from the experiments it has been shown that SiGe HBTs can tolerate up to 200 Mrad dose \cite{Cressler_Extreme}.

For four different SiGe generations, Fig.~\ref{fig:Unity_gain_cutoff_frequency_4th_gen_SiGe_HBT} explains the cutoff frequency vs. the bias current density, before and after the exposure of the same total dose as earlier, the cutoff frequencies remain very stable despite high radiation exposure. These characteristics make SiGe HBTs distinctive among the available semiconductor technologies, and very suitable for any extraterrestrial application. This TID tolerance of SiGe HBTs comes from the device structure that the damaged interface is physically isolated from the carrier transport path \cite{Jeff}. 
%the base region of an HBT that has embedded a strained SiGe alloy to speed up the carrier movement to ultimately speed up the HBT, not from the Ge doping \cite{Cressler_Extreme}.

\begin{figure} 
         \centering 
         \includegraphics[scale=0.4500]{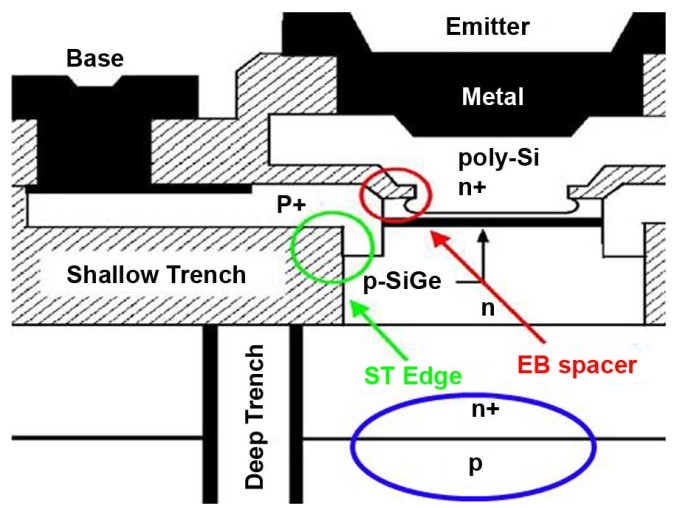}
         \caption[The vulnerable points affiliated with the emitter-base (EB) spacer, the shallow trench isolation  (STI) edge, and the $n^+$-$p^_$ sub-collector-to-substrate junction in a SiGe HBT's schematic cross-section.]{The vulnerable points affiliated with the emitter-base (EB) spacer and the shallow trench isolation  (STI) edge, and the $n^+$-$p^-$ sub-collector-to-substrate junction in a SiGe HBT's schematic cross-section \cite{Cressler_Extreme}.}
         \label{fig:Cross_section_of_Commercial_SiGe_HBT_damage}
\end{figure}

\begin{figure} 
         \centering 
         \includegraphics[scale=0.450]{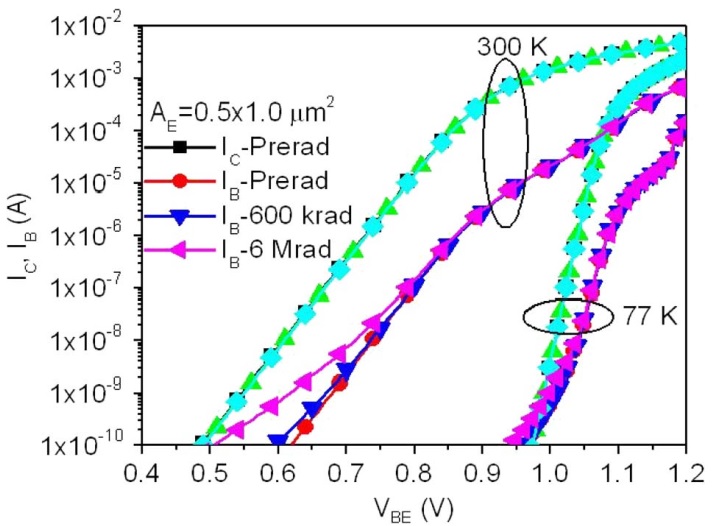}
        \caption[Pre and post-radiation, I-V characteristics of a 1st-gen SiGe HBT at temperatures 300K and 77K, when subject to space-relevant 63-MeV protons with total dose up to multi-Mrad.]{Pre and post-radiation, I-V characteristics of a 1st-gen SiGe HBT at temperatures 300K and 77K, when subject to space-relevant 63-MeV protons with total dose up to multi-Mrad \cite{Cressler_Extreme}.}
        \label{fig:IV_Char_1stgen_SiGe_HBT_63MeV_proton}
\end{figure}

As shown in  Fig.~\ref{fig:Cross_section_of_Commercial_SiGe_HBT_damage}, the radiation tolerance of a BJT usually gets weakened because of the charge accumulation in the oxide of the emitter–base (EB) spacer but for SiGe HBT three structural features work for its side, like a very thin (e.g., $\leq$100 nm) EB spacer and situated in the epitaxial base with very high doping, the base is also extremely narrow (e.g., $\leq$100 nm) and heavily doped, and the transistor core transport path is far away from the thin shallow trench isolation  (STI) in the CB junction \cite{Cressler_Extreme}. Fig.~\ref{fig:IV_Char_1stgen_SiGe_HBT_63MeV_proton} shows the environmental invariance of the SiGe electronics with its built-in TID immunity works even better if they are operated in cryogenic temperatures. This shows SiGe HBTs are wide-temperature and radiation-invariant transistors \cite{Prakash}.

\section{Why PMOS+SiGe HBT circuit design approach?}
\label{Sec:whypmos+hbt}

%\subsection{Performence of NMOS in Cold and Radiation}
\subsection{Performance of NMOS in Cold and Radiation}

From the above discussion, NMOS has the following issues in extreme temperature (cryogenic) and radiation:

\begin{enumerate}
    \item The projected lifetime for NMOS is ~2 orders of magnitude shorter than PMOS in cryogenic temperature due to the hot carrier effect (HCE) \cite{Brookhavenppt, Shaorui}.
    \item Due to TID, NMOS will have positive charge accumulation in the oxide layer, and over time it will become leaky and harder to turn off \cite{Moscap, TIDMOSFETTI}.
    \item For NMOS, multiple leakage paths get activated after the radiation exposure \cite{Moscap, TIDMOSFETTI}. 
    \item The leakage current in NMOS varies up to 6 orders of under radiation whereas PMOS has near zero leakage current after the radiation exposure due to positive charge accumulation in the oxide \cite{Moscap}. 
    \item The annular gate structure greatly enhances TID immunity for the NMOS device. However, most commercial CMOS foundries do not support this type of NMOS device structure \cite{9hp_manual}. 
\end{enumerate}

\subsection{Performance of PMOS+SiGe HBTs in Cold and Radiation}

The performance characteristics of the PMOS in extreme temperature (cryogenic) and radiation:

\begin{enumerate}
    \item The projected lifetime for PMOS is ~2 orders of magnitude longer than NMOS in cryogenic temperature even though the hot carrier effect (HCE) is present in PMOS \cite{Brookhavenppt, Shaorui}.
    \item Due to TID, PMOS will have similar positive charge accumulation in the oxide layer,  it will have zero leakage current as the positive charge will attract more electrons in the channel and make it harder to invert. So, over time PMOS will be harder to turn on \cite{Moscap, TIDMOSFETTI}.
    \item PMOS has a limited number of leakage paths that get activated after the radiation exposure than NMOS \cite{Moscap, TIDMOSFETTI}. 
    
\end{enumerate}

 The summary of the performance characteristics of the SiGe HBT in extreme temperature (cryogenic) and radiation:

\begin{enumerate}
    \item SiGe HBT can tolerate TID up to some multiMrad levels because of its internal built-in TID tolerance with dependable performance in the multiple generations of the scaling node \cite{Cressler_Extreme, Babcock, Roldan, Zhang, Proton_tolerance, Sutton, Sutton1}. SiGe HBTs can tolerate up to 200 Mrad dose \cite{Cressler_Extreme}. 
    \item  TID tolerance of SiGe HBTs comes from the device structure that the damaged interface is physically isolated from the carrier transport path \cite{Jeff}.
    \item SiGe electronics with their built-in TID immunity work even better if they are operated in cryogenic temperatures (up to 4.5K) \cite{Prakash}. 
\end{enumerate}

%\subsection{Performence of PMOS+HBT in Cold and Radiation}

\subsection{Why PMOS+SiGe HBT is needed ?}

The need for extreme environment electronics is growing exponentially as satellite technology, communications infrastructure around the world, space exploration, particle physics detectors, automotive electronics inside the engine, weapon systems, sensor systems, and health imaging systems, remote imaging systems. This harsh environment creates ``reliability issues'' for conventional electronics, i.e. commercial-off-the-shelf (COTS) electronic components, e.g. Si MOSFET as part of a satellite system may fail very quickly by exposing it to cold temperatures and ionizing radiation in space. To solve this problem, all electronics need to be hardened based on their utilization in a target environment by controlling the flow of fabrication ('hardening by process') and / or changing different topologies of circuits and / or system architecture ('hardening by design') \cite{Cressler_Extreme}. For harsh environmental conditions like Mars's surface where the daily temperature swings between  {-$135^{\circ}$}{C} to  {+$90^{\circ}$}{C} from Martian night to Martian day, NASA housed all of their conventional COTS electronics for sensing, controlling motor/actuators, and communication circuits in a ``warm-electronics-box'' for Mars exploration missions to protect them from cold and radiation \cite{YChen}. But this is a very crude solution for advanced robotics and rovers for space exploration, where all electronics are exposed to the environment because of their placement at 'the point of use' or close to the loads while collecting samples or sensing the overall health of the rover for scientific exploration or to respond to environmental input from the sensors. Usually, these Martian or Lunar rovers are built with intelligent electronic nodes, which are distributed over the remote part of the vehicle, which makes it very inefficient to protect those nodes from the harsh environment with the protective 'warm boxes'. Due to the use of protective electronic 'warm boxes', the electronic systems on the rovers are not very distributed, bulky, non-modular, and less reliable \cite{Cressler_Extreme}. 

\section{Conclusion}\label{section:Conclusions}

Now, the solution to these extreme environmental issues requires environmentally invariant electronics (i.e. robust operation regardless of the environment without any protection). Mission designers and vehicle architects can leverage these environmentally invariant electronics for any remote probe, manned missions, or even Martian or lunar colonization. Silicon germanium (SiGe) technology offers a significant potential for environmentally invariant electronic technology, which shows remarkable device characteristics (SiGe HBT) in a radiation environment at both cryogenic and elevated temperatures \cite{Cressler_Extreme}. The modern SiGe technology offers a BiCMOS platform (SiGe HBT + Si CMOS), but the reliability of the CMOS device is still a concern, especially the performance issues in NMOS in extreme environments, as mentioned above. The PMOS+HBT circuit design approach gives circuit designers a chance to utilize the best characteristics of PMOS, polysilicon resistors, and SiGe HBTs in extreme cold and radiation environments.

\end{document}